\documentclass[twocolumn,superscriptaddress,showpacs,amsmath,amssymb,aps]{revtex4}
\usepackage{graphicx}

\begin{document}

\title{Markov Chain Monte Carlo Method without Detailed Balance}

\author{Hidemaro Suwa}
\affiliation{Department of Applied Physics, University of Tokyo, Tokyo 113-8656, Japan}
\author{Synge Todo}
\affiliation{Department of Applied Physics, University of Tokyo, Tokyo 113-8656, Japan}
\affiliation{CREST, Japan Science and Technology Agency, Kawaguchi 332-0012, Japan}

\date{\today}

\begin{abstract}
We present a specific algorithm that generally satisfies the balance condition without imposing the
detailed balance in the Markov chain Monte Carlo. In our algorithm, the average rejection rate is
minimized, and even reduced to zero in many relevant cases. The absence of the detailed balance also
introduces a net stochastic flow in a configuration space, which further boosts up the convergence. We
demonstrate that the autocorrelation time of the Potts model becomes more than 6 times shorter than that
by the conventional Metropolis algorithm. Based on the same concept, a bounce-free worm algorithm for
generic quantum spin models is formulated as well.
\end{abstract}

\pacs{02.70.Tt, 05.10.Ln, 02.50.-r, 02.70.Ss}

\maketitle

The Markov chain Monte Carlo (MCMC) method, which is a vital tool for investigating almost all kinds of statistical problems, especially systems with multiple degrees of freedom, is being applied extensively across the various disciplines, such as statistics, economics, and bioinformatics, not to mention physics~\cite{LandauB2005,GilksRS1996,RobertC2004}.
In the MCMC method for a stationary distribution, the balance condition [BC, Eq.~(\ref{eqn:bc}) below] and the ergodicity must be imposed; the former ensures the invariance of a target distribution, while the latter does the convergence to the invariant~\cite{Tierney1994,MeynT1993}.
Although an MCMC method satisfying these two conditions guarantees unbiased results in infinite time in principle, a rapid relaxation, that is, a short correlation time, is essential for the method to work in practice.  The longer the autocorrelation time is, the larger the error bar becomes.
In order to achieve weaker correlations in a Markov sequence, a number of variants have been invented so far, e.g., the Swendsen-Wang algorithm~\cite{SwendsenW1987} and the loop algorithm~\cite{EvertzLM1993}.
The extended ensemble methods, such as the multicanonical method~\cite{BergN1992} and the exchange Monte Carlo method~\cite{HukushimaN1996}, have also been proposed and applied successfully to protein folding problems, spin glasses, etc.

In most practical implementations of the MCMC method, the detailed
balance condition (DBC), the reversibility, is imposed,
where every elementary transition is forced to balance with a
corresponding inverse process.  Thanks to the DBC, it becomes easy
to find qualified transition probabilities.  In the
meantime, it has long been considered difficult to satisfy the BC without
imposing the DBC, and attempts to reduce autocorrelations in
Markov sequences have concentrated on optimizing transition
probabilities within the
DBC~\cite{Liu1996,PolletRVH2004}.  Here, we need to be reminded that the DBC is
not a necessary condition for the invariance.  
The BC, which is a weaker condition than the DBC, is
mathematically sufficient~\cite{Tierney1994,MeynT1993}.  In fact,
the DBC is often broken secretly, even though the DBC
is used apparently to define the transition probabilities.
The single spin update in a classical
system is such an example.
The random
update, where a spin to be flipped is
chosen uniformly randomly among all spins, satisfies the DBC strictly.  On the other hand, the DBC
is broken in the sequential update, where spins are swept in a fixed
order.  The DBC is satisfied only locally, that is, only in each
spin flip, and the BC is eventually fulfilled in one
sweep~\cite{ManousiouthakisD1999}.

In this Letter, we present a simple and versatile algorithm to find a
set of transition probabilities in the MCMC method, which fulfills the BC but
breaks the DBC even locally.  As the BC is a weaker condition than the
DBC, the solution space of transition probabilities is
enlarged, and rejection rates can be reduced as the result.  We show
that by the present algorithm the average rejection rate is indeed
minimized, and even reduced to zero in many relevant cases.
Furthermore, breaking the DBC introduces a net stochastic flow
in the configuration space.  It will boost up the relaxation further
by suppressing random walk
behavior~\cite{Adler1981,DuaneKPR1987,DiaconisHN2000}\footnote{The
  overrelaxation method~\cite{Adler1981} and the hybrid Monte
  Carlo~\cite{DuaneKPR1987} both satisfy the DBC, mistakenly believed
  to break it.}.

In what follows, after describing our specific algorithm for finding
a rejection-minimized solution, we will demonstrate its effectiveness
in the single spin update of the Potts model~\cite{Wu1982} and the
worm (directed-loop) update~\cite{ProkovievST1998,SyljuasenS2002}
of the quantum Heisenberg spin chain in a magnetic field.  In both cases, it
is established that our algorithm boosts up the relaxation significantly
in comparison with the conventional algorithms, such as the Metropolis
(also called Metropolis-Hastings) algorithm~\cite{MetropolisRRTT1953,
  Hastings1970}, or the heat bath algorithm (Gibbs
sampler)~\cite{Creutz1980,GemanG1984}.
Especially, a bounce-free (rejection-free) worm algorithm can be formulated for generic quantum spin models, by which the autocorrelation time is often reduced by orders of magnitude as we will see below.

We start with considering an elementary update process,
e.g., flipping a single spin in the Ising or Potts models, or choosing an
exit at an operator in the worm algorithm (see below).  Given an 
environmental configuration, we would have $n$ candidates (including
the current one) for the next configuration.  The weight of each
candidate configuration (or state) is given by $w_i$ ($i=1,\/ ...,
n$), to which the equilibrium probability distribution is
proportional.  Although in the formulation of the MCMC method, the BC (or the
DBC) is usually expressed in terms of the weights $\{w_i\}$ and the
transition probabilities $\{p_{i \rightarrow j}\}$, it is more
convenient to introduce a quantity $v_{i \rightarrow j} := w_i p_{i
  \rightarrow j}$, which corresponds to the amount of (raw) stochastic
flow from state $i$ to $j$.  The law of probability conservation and
the BC are then expressed as
\begin{align}
  w_{i} & = \sum_{j=1}^n v_{i \rightarrow j} \qquad \forall \, i
  \label{eqn:conservation}
  \\
  w_{j} &= \sum_{i=1}^n v_{i \rightarrow j} \qquad \forall \, j,
  \label{eqn:bc}
\end{align}
respectively.  The average rejection rate is written as $\sum_i v_{i \rightarrow
  i} / \sum_i w_i$.  Also, it is straightforward to confirm that $\{v_{i
  \rightarrow j}\}$ satisfy
\begin{equation}
  v_{i \rightarrow j} = \frac{1}{n-1} \min( w_i, w_j ) \qquad i \neq j
  \label{eqn:metropolis}
\end{equation}
for the Metropolis algorithm, and
\begin{equation}
  v_{i \rightarrow j} = \frac{w_i w_j}{\sum_{k=1}^n w_k} \qquad \forall \, i, \, j
  \label{eqn:heatbath}
\end{equation}
for the heat bath algorithm, where the detailed balance, i.e., the absence of net
stochastic flow, is manifested by the symmetry under the interchange
of the indices: $v_{i \rightarrow j} = v_{j \rightarrow i}$.

\begin{figure}
\includegraphics[width=5.9cm]{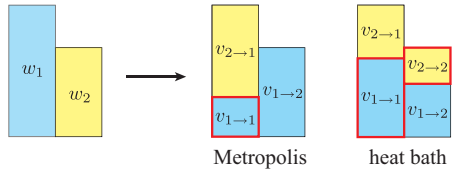}
\caption{Example of weight landfill by the Metropolis and heat bath algorithms
  for $n=2$.  The amount of landfilled weight, $v_{i \rightarrow j}$, is defined
  by Eqs.~(\ref{eqn:metropolis}) and (\ref{eqn:heatbath}),
  respectively.  The regions with thick frame denote the rejection rates.}
\label{fig:landfill-DBC}
\end{figure}

Our task is to find a set $\{v_{i \rightarrow j}\}$ that minimizes 
the average rejection rate while satisfying
Eqs.~(\ref{eqn:conservation}) and (\ref{eqn:bc}).  This procedure can
be understood visually as weight landfill, where we move (or
allocate) some amount of weight ($v_{i \rightarrow j}$) from state $i$
to $j$ keeping the entire shape of the weight boxes intact.  For
catching on this landfill picture, let us think at first the case 
with $n=2$ as in the single
spin update of the Ising model.  Figure~\ref{fig:landfill-DBC} shows the
landfill when the Metropolis and heat bath algorithms
[Eqs.~(\ref{eqn:metropolis}) and (\ref{eqn:heatbath})] are applied,
where the average rejection rate ($\propto v_{1 \rightarrow 1} + v_{2
  \rightarrow 2}$) clearly remains finite.  Indeed, for $n=2$ the
Metropolis algorithm gives the best solution, i.e., the minimum
average rejection rate even within the BC [see
Eq.~(\ref{eq:rejection}) below].

\begin{figure}
\includegraphics[width=7.9cm]{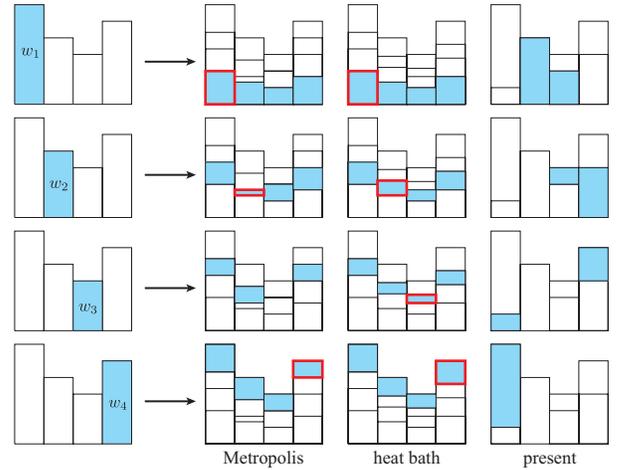}
\caption{Example of weight landfill by the Metropolis, heat
  bath, and the proposed algorithms for $n=4$. In the present algorithm, first
  the maximum weight ($w_1$) is allocated to the second box.  It
  saturates the second box, and the remainder is all put into the
  third one (first row).  Next, $w_2$ is allocated to the partially
  filled box and the subsequent box (second row).  The same
  procedure is repeated for $w_3$ and $w_4$.
  This proposed algorithm is rejection free, while there remain
  finite rejection rates in the others as indicated by the thick frames.}
\label{fig:landfill-BC}
\end{figure}

For $n \ge 3$, on the other hand, we can get ahead by breaking with the
DBC.  In Fig.~\ref{fig:landfill-BC}, we show examples of weight
landfill in the case with $n=4$ by using the Metropolis and the heat
bath algorithms together with a solution that does not satisfy the
DBC.  The first two again have finite rejection rates, while the last
is rejection free (i.e., $v_{i \rightarrow i}=0$ $\forall\,i$).
Although a solution is not unique obviously, we propose the
following procedure as a concrete algorithm to find a solution for
general $n$. (i) Choose a configuration with maximum weight.  If two or
more configurations have the same maximum weight, choose one of them.
In the following, we assume $w_1$ is the maximum without loss of
generality.  The order of the remaining weights does not matter.  (ii)
Allocate the maximum weight $w_1$ to the next box ($i=2$).  If the
weight still remains after saturating the box, reallocate the remainder to the next
($i=3$).  Continue until the weight is all allocated.  (iii) Allocate
the weight of the first landfilled box ($w_2$) to the last partially
filled box in step ii).  Continue the allocation likewise.  iv) Repeat
step iii) for $w_3$, $w_4,\/ ..., w_n$.  Once all the boxes with $i \ge 2$
are saturated, landfill the first box ($i=1$) afterward.

In the above procedure, all the boxes are landfilled without any
space; that is, it satisfies the BC [Eq.~(\ref{eqn:bc})].  Since the BC
is satisfied in each elementary transition, it is fulfilled in one
sweep as well.  It is also clear that the second and subsequent boxes must be
already saturated when the allocation of its own weight is
initiated, since $w_1$ is the maximum.  By this procedure, $\{ v_{i
  \rightarrow j} \}$ are determined as
\begin{equation}
v_{i \rightarrow j } = \max( 0, \, \min( \Delta_{ij}, \, w_i + w_j - \Delta_{ij}, \, w_i, \, w_j ) ),
\label{eq:BC-expression}
\end{equation}
where
\begin{align}
\Delta_{ij} &:= S_i - S_{j-1} + w_1 \qquad 1 \leq i, \, j \, \leq n \\
S_i &:= \sum_{k=1}^i w_k \qquad 1 \leq i \leq n \\
S_0 &:= S_n.
\end{align}
Especially, from Eq.~(\ref{eq:BC-expression}) we obtain
\begin{equation}
v_{i \rightarrow i} = \begin{cases}
  \max( 0, \, 2 w_1 - S_n ) & i = 1 \\
  0 & i \geq 2.
\end{cases}
\label{eq:rejection}
\end{equation}
That is, a rejection-free solution can be obtained, if
\begin{eqnarray}
w_1 \leq \frac{S_n}{2} \equiv \frac{1}{2}\sum_{k=1}^n w_k
\label{eq:rejection-free}
\end{eqnarray}
is satisfied.  On the contrary, when inequality
(\ref{eq:rejection-free}) is not satisfied, one has to necessarily
assign the maximum weight to itself, as it is bigger than the sum of
the rest.  Thus, the present solution is optimal within the BC,
in the sense that it minimizes the average rejection rate.

\begin{figure}
\includegraphics[width=7.0cm]{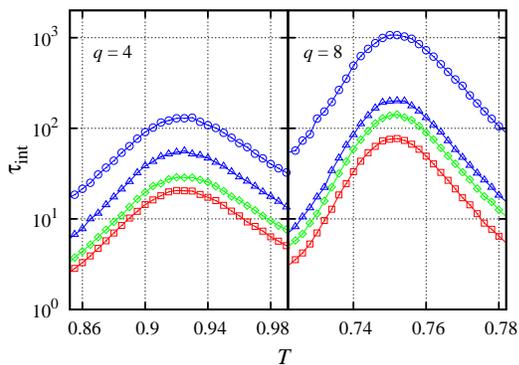}
\caption{Autocorrelation time of the square of order parameter near the transition temperature ($T \simeq 0.910$ and 0.745, respectively) in the 4-state (left) and 8-state (right) Potts models by the Metropolis (circles), heat bath (triangles), LOU (diamonds), and present (squares) methods.  The system size is $16 \times 16$. 
The error bars are the same order with the point sizes.}
\label{fig:Potts-stf}
\end{figure}

We close the introduction of our algorithm with a note
about the ergodicity. It is far from trivial to prove that the
present algorithm, as well as the Metropolis algorithm,
satisfies the ergodicity, since many of the transition probabilities
become zero exactly. In principle, however, one
can always ensure the ergodicity by combining the present
algorithm with the heat bath method. Another way to
ensure the ergodicity is randomly choosing one of the
sets of transition probabilities obtained by different landfill
order, though we have not observed any glimpse of ergodicity
breaking in the following simulations.

In order to assess the
effectiveness of the present algorithm, we investigate the
autocorrelations in the ferromagnetic $q$-state Potts models on the
square lattice~\cite{Wu1982}, which exhibit a continuous
($q \le 4$) or first-order ($q>4$) phase transition at $T=1 / \ln
(1+\sqrt{q})$.  We calculate the autocorrelation time of the square of
order parameter for $q=4$ and 8 by several algorithms.  The
autocorrelation time $\tau_{\rm int}$ is estimated through the
relation: $ \sigma^2 = ( 1 + 2 \tau_{\rm int} )
\sigma_{0}^2$, where $\sigma_0^2$ is the variance of the raw
time series data and $\sigma^2$ is the true variance calculated
from the binned data using a bin size much larger than the $\tau_{\rm
  int}$~\cite{LandauB2005}.
In Fig.~\ref{fig:Potts-stf},
it is clearly seen that our algorithm significantly boosts up the
relaxation in both models in comparison with the conventional methods.
In the 4-state Potts model, the autocorrelation time becomes nearly 6.4
times shorter than that by the Metropolis algorithm, 2.7 times than the heat
bath algorithm, and even 1.4 times than the locally optimal update
(LOU) by Pollet {\it et al}.~\cite{PolletRVH2004}, which is considered
as one of the best solutions within the DBC.  Furthermore, the present
algorithm is increasingly advantaged as $q$ increases.

\begin{figure}
\includegraphics[width=8.6cm]{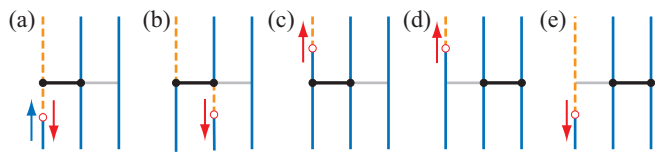}
\caption{Extension of the worm-going pathway in the $S=1/2$ model.
  Here, dashed (solid) vertical lines denote spin up (down).  The head
  of worm (open circle) moves on the worldline~(a), and scatters at
  the operator (horizontal thick line).
   As candidate configurations, we introduce operator-flip updates (d)--(e)
  in addition to the conventional ones (a)--(c).  Note
  that in (e) the position of the operator is shifted simultaneously in
  contrast to the simple bounce process
  (a).
}
\label{fig:flip}
\end{figure}

Next, we move onto the quantum Monte Carlo methods.  The worm algorithm for quantum spin and lattice
boson models is formulated based on either the Euclidean pathintegral or the high-temperature
series~\cite{ProkovievST1998,SyljuasenS2002}.  One Monte Carlo
sweep of the worm algorithm consists of the diagonal update, where
operators are inserted or removed without changing the shape of
worldlines, and the off-diagonal update, where the worldlines (and
the type of operators) are updated with keeping the position of
operators unchanged.  In the latter process, a pair of creation and
annihilation operators, which is called a worm, is inserted on a
worldline (pair creation), and one of them (called the head) is moved
stochastically until the head and the tail destroy each other (pair
annihilation).  As a thorny problem, a bounce process, where the head
just backtracks and cancels the last update, has been generally
inevitable within the DBC.  Here, as an example, we consider the
$S=1/2$ antiferromagnetic XXZ model:
\begin{equation}
{\cal H} = \!\!\sum_{<i, j>} \!\!\left( S_i^x S_{j}^x + S_i^y S_j^y + \Delta S_i^z S_j^z - C \right) - h \sum_i S_i^z,
\end{equation}
where we introduce an arbitrary parameter $C$ controlling the ratio
between the diagonal and off-diagonal weights.  In the head scattering
process at an operator, only three among four exits have
a nonzero weight due to the conservation of the total $S^z$ [(a)--(c) in
Fig.~\ref{fig:flip}].  At the Heisenberg point ($\Delta=1$), there
remain finite bounce probabilities except at $h=0$ within the
DBC~\cite{SyljuasenS2002,PolletRVH2004,AletWT2005}.  Unfortunately,
the situation does not improve much even in the BC, because the total number of
candidates is too small.  However, the condition~(\ref{eq:rejection-free})
provides us a clear prospect; by increasing the number of
candidates, a bounce-free algorithm will be realized.  According to this
strategy, we introduce {\it operator-flip} updates, where
sites on which an operator acts are shifted simultaneously
(Fig.~\ref{fig:flip}).  By the operator flip together with
the constant $C$ chosen as
\begin{equation}
  C = \max\Big( \frac{1}{4}( 2 \Delta + 3h - 1 ), \, \frac{1}{8}( \Delta + 3h + 1 ) \Big), \label{eq:bounce-free}
\end{equation}
we can actually eliminate the bounce process absolutely.

\begin{figure}
\includegraphics[width=7.0cm]{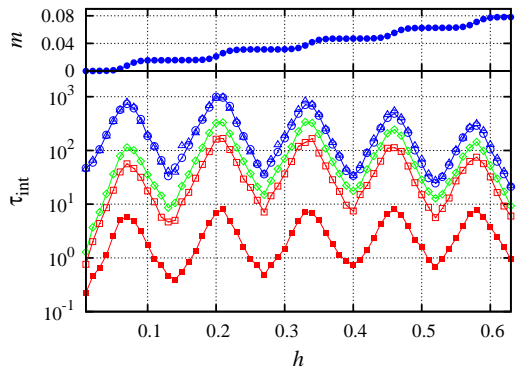}
\caption{Magnetic field dependence of magnetization (upper) and autocorrelation time (lower) of the $S=1/2$ antiferromagnetic Heisenberg chain ($L=64$, $T=1/2L$).  The maximum autocorrelation time is $1.0 \times 10^3$ by the worm update with Metropolis (circles), $9.8 \times 10^2$ by the worm with heat bath (triangles), $3.3 \times 10^2$ by the LOU (diamonds), an improvement of the directed loop, and $1.7 \times 10^2$ by the present algorithm (squares).  By the bounce-free worm with the operator flip, $\tau_{\rm int}$ is further reduced down to 8.1 (solid squares).}
\label{fig:Heisenberg}
\end{figure}

The autocorrelation data of the magnetization in the Heisenberg chain
($\Delta=1$) are shown in Fig.~\ref{fig:Heisenberg}.  Amazingly, the
bounce-free worm algorithm with the operator flip is faster by about 2
orders of magnitude than the Metropolis and the heat bath algorithms.
Also in high-$S$ spin systems, the bounce-free worms can be
constructed by representing the partition function by decomposed
$S=1/2$ spins~\cite{TodoK2001}.  Our idea of breaking the DBC and
operator-flip updates are also applied to general bosonic models
effectively, because bosonic worms get bounce-minimized with more
candidates.

In the present study, we have developed a simple and versatile MCMC
algorithm that generally satisfies the BC without imposing the DBC. In
our algorithm, the average rejection rate gets minimized, which
reduces the autocorrelation time significantly in comparison with the
conventional methods based on the DBC.  We also have introduced
operator-flip updates in the worm algorithm, yielding a bounce-free
algorithm in generic spin models.  The present concept can be
naturally extended to systems with continuous state variables by
replacing the weight landfill presented here with an asymmetric random
cyclic shifting in a cumulative probability distribution function.
Our approach beyond the DBC can be universally applied to any MCMC
methods, even to ingenious established methods, such as the cluster
algorithms and the extended ensemble methods, and will undoubtedly
improve the relaxation in Markov sequences.

Most simulations were performed on T2K Supercomputer at University of
Tsukuba.  The program was developed based on the ALPS
library~\cite{TodoK2001,ALPS2007}.  We acknowledge support by Grant-in-Aid for
Scientific Research Program (No.\,20540364) from JSPS, and by Grand
Challenges in Next-Generation Integrated Nanoscience, Next-Generation
Supercomputer Project from MEXT, Japan.

\bibliography{bc}

\end{document}